\documentclass[useAMS,usenatbib,psfig]{mn2e}
\usepackage{graphicx}
\begin{document}
\title[Optical evolution of Nova Ophiuchi 2007 = V2615 Oph]{Optical evolution of Nova Ophiuchi 2007 = V2615 Oph}
\author[U. Munari et al.]{U. Munari$^{1}$, A. Henden$^{2}$,
M. Valentini$^{1,3}$, A. Siviero$^{1}$, S. Dallaporta$^{4}$,
P. Ochner$^{4}$ and S.Tomasoni$^{4}$\\
$^{1}$INAF Astronomical Observatory of Padova, 36012 Asiago (VI), Italy\\
$^{2}$AAVSO, 49 Bay State Road, Cambridge, MA 02138, USA\\
$^{3}$Isaac Newton Group of Telescopes, Apartado de Correos 321, E-38700 Santa Cruz de La Palma, Spain\\
$^{4}$ANS Collaboration, c/o Osservatorio Astronomico, via dell'Osservatorio 8, 36012 Asiago (VI), Italy}

\date{Accepted .... Received ....; in original form ....}

\pagerange{\pageref{firstpage}--\pageref{lastpage}} \pubyear{2008}
 
\maketitle

\label{firstpage}

\begin{abstract}
The moderately fast Nova Oph 2007 reached maximum brightness on March 28,
2007 at $V$=8.52, $B-V$=+1.12, $V-R_{\rm C}$=+0.76, $V-I_{\rm C}$=+1.59
and $R_{\rm C}-I_{\rm C}$=+0.83, after fast initial rise and a pre-maximum
halt lasting a week.  Decline times were $t^{V}_{2}$=26.5, $t^{B}_{2}$=30,
$t^{V}_{3}$=48.5 and $t^{B}_{3}$=56.5~days. The distance to the nova is
$d$=3.7 $\pm$0.2 kpc, the height above the galactic plane $z$=215~pc, 
the reddening $E_{B-V}$=0.90 and the absolute magnitude at maximum 
$M_V^{\rm max}$=$-$7.2 and $M_B^{\rm max}$=$-$7.0. The spectrum four days
before maximum resembled a F6 super-giant, in agreement with broad-band
colors. It later developed into that of a standard 'FeII'-class nova. Nine
days past maximum, the expansion velocity estimated from the width of
H$\alpha$ emission component was $\sim$730~km/s, and the displacement from
it of the principal and diffuse enhanced absorption systems were $\sim$650 and
1380 km/s, respectively. Dust probably formed and disappeared during the period
from 82 to 100 days past maximum, causing (at peak dust concentration) an
extinction of $\Delta B$=1.8 mag and an extra $\Delta E_{B-V}$=0.44
reddening.
\end{abstract}

\begin{keywords}
stars: classical novae
\end{keywords}

\section{Introduction}

Nova Oph 2007 (= V2615 Oph, hereafter NOph07) was discovered by H. Nishimura
at $\sim$10~mag on photographic film exposed on Mar 19.81 UT (cf. Nakano
2007), and confirmed spectroscopically by Naito and Narusawa (2007).  Das et
al. (2007) reported infrared spectroscopy showing unusual and strong CO
molecular bands in emission around optical maximum. An early report on
pre-maximum spectral appearance of NOph07 in the optical was provided by
Munari et al. (2007). Rudy et al. (2007) announced dust condensation
occurring in NOph07 during May 2007, while Henden and Munari (2007)
calibrated a $B$$V$$R_{\rm C}$$I_{\rm C}$ photometric comparison sequence
and measured an accurate astrometric position ($\alpha_{\rm J2000}$ = 17 
42 44.013~$\pm$0$^{\prime\prime}$.03, $\delta_{\rm J2000}$ = $-$23 40 
35.05~ $\pm$0$^{\prime\prime}$.07).

\section{Observations}

The $B$$V$$R_{\rm C}$$I_{\rm C}$ photometric evolution of NOph07 has been
monitored, for seven months and over a seven magnitude decline, with three
different telescopes (identified by $a$,$b$,$c$ letters below and in Figures
1 and 2):
($a$) the Sonoita Research Observatory (SRO) 0.35-m Celestron C14 robotic
telescope using $B$$V$$R_{\rm C}$$I_{\rm C}$ Optec filters and an SBIG
STL-1001E CCD camera, 1024$\times$1024 array, 24 $\mu$m pixels $\equiv$
1.25$^{\prime\prime}$/pix, with a field of view of
20$^\prime$$\times$20$^\prime$; ($b$) the 0.30-m Meade RCX-400 f/8
Schmidt-Cassegrain telescope owned by Associazione Astrofili Valle di Cembra
(Trento, Italy). The CCD is a SBIG ST-9, 512$\times$512 array, 20 $\mu$m
pixels $\equiv$ 1.72$^{\prime\prime}$/pix, with a field of view of
13$^\prime$$\times$13$^\prime$. The $B$ filter is from Omega and the
$V$$R_{\rm C}$$I_{\rm C}$ filters from Custom Scientific; ($c$) the 0.50-m
f/8 Ritchey-Cretien telescope operated on top of Mt. Zugna by Museo Civico
di Rovereto (Trento, Italy) and equipped with Optec $U$$B$$V$$R_{\rm
C}$$I_{\rm C}$ filters. The CCD is an Apogee Alta U42 2048$\times$2048
array, 13.5 $\mu$m pixels $\equiv$ 0.70$^{\prime\prime}$/pix, with a field
of view of 24$^\prime$$\times$24$^\prime$. The overall $B$$V$$R_{\rm
C}$$I_{\rm C}$  light- and color-curves are presented in Figure~1.

All photometric measurements were carefully tied to the $B$$V$$R_{\rm
C}$$I_{\rm C}$ calibration sequence of Henden and Munari (2007). They are
listed in Table~1 (available electronic only). In all, we obtained 442
independent photometric measures (115 in $V$, 106 in $B-V$, 114 in $V-R_{\rm
C}$, 106 in $R_{\rm C}-I_{\rm C}$, 107 in $V-I_{\rm C}$) distributed over 67
different nights. The mean poissonian errors of the photometric points in
Figure~1 is 0.004~mag in $V$, 0.006 in $B-V$, 0.006 in $V-R_{\rm C}$, 0.003
in $R_{\rm C}-I_{\rm C}$ and 0.005 in $V-I_{\rm C}$. The mean r.m.s. of
standard stars from the linear fit to color equations is 0.022~mag in $V$,
0.032 in $B-V$, 0.029 in $V-R_{\rm C}$, 0.018 in $R_{\rm C}-I_{\rm C}$ and
0.043 in $V-I_{\rm C}$. In spite of the excellent color transformations of
all three instruments to the Henden and Munari (2007) comparison sequence,
the presence of strong emission lines in the spectrum of NOph07 causes
unavoidable differences between nova data recorded with different
telescopes, at the level of a few hundreds of a magnitude as Figure~2 well
illustrates.

Low and medium resolution, absolutely fluxed spectra of NOph07 were obtained
on March 22.17 and 24.16 UT, 2007 with the AFOSC imager+spectrograph mounted on the
1.82m telescope operated in Asiago by INAF Astronomical Observatory of
Padova. It is equipped with a Tektronix TK1024 thinned CCD, 1024$\times$1024
array, 24 $\mu$m pixel, with a scale perpendicular to dispersion of
0.67~$^{\prime\prime}$/pix. All observations were performed with a
1.26$^{\prime\prime}$ slit aligned along the parallactic angle. We adopted a 300 ln/mm
grism, covering the 3500-7780~\AA\ interval at 4.24~\AA/pix dispersion, and
a 1720 ln/mm volume phase holographic grism, for the range 6400-7050~\AA\ at
0.64~\AA/pix.

   \begin{figure}
     \centering
     \includegraphics[width=8.5cm]{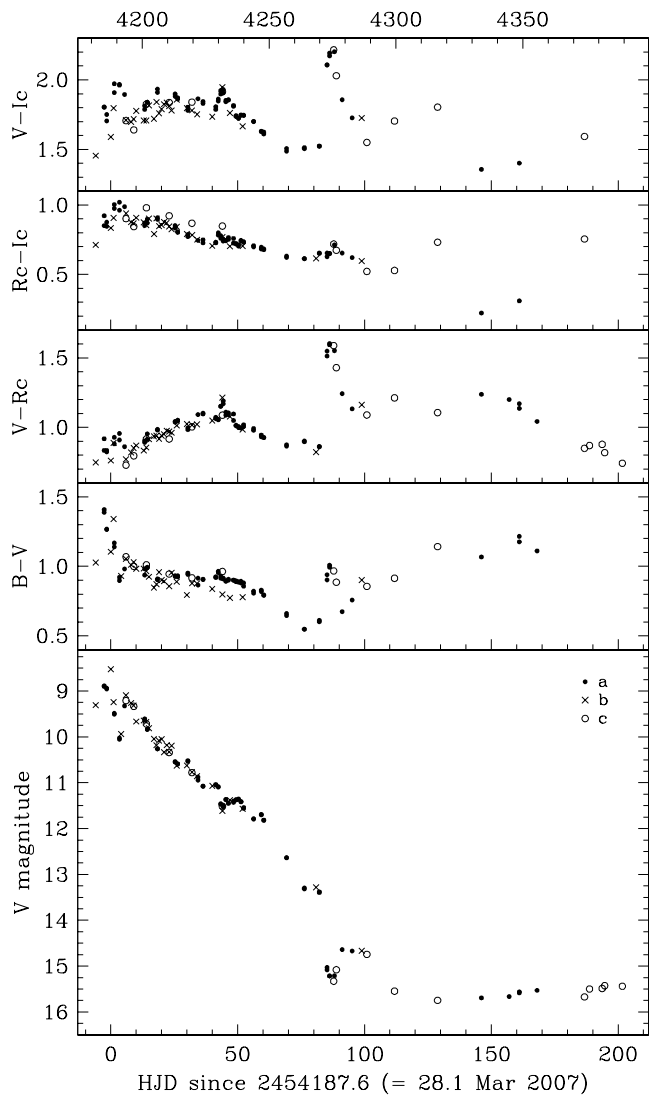}
   \caption{Light- and color-curves of Nova Oph 2007. The different symbols
            identify the telescopes used to collect the data which are 
            detailed in Sect. 2. The ordinates at bottom are days from
            maximum brightness, those at top are heliocentric JD minus
            2450000.}
   \end{figure}

Low and medium resolution, absolutely fluxed spectra of NOph07 were obtained
on April 6, 2007 also with the 0.6-m telescope of Osservatorio Astronomico
G. Schiaparelli (Varese, Italy), equipped with a grating spectrograph and
SBIG ST-10XME CCD, 2184$\times$1472 array, 6.8 $\mu$m pixel. The slit width
was 2.0$^{\prime\prime}$. A 600 ln/mm grating was used to cover the
3900-7100~\AA\ range at 1.76~\AA/pix, and a 1800 ln/mm grating for the
6200-6900~\AA\ range at 0.32~\AA/pix.

   \begin{figure}
     \centering
     \includegraphics[width=8.5cm]{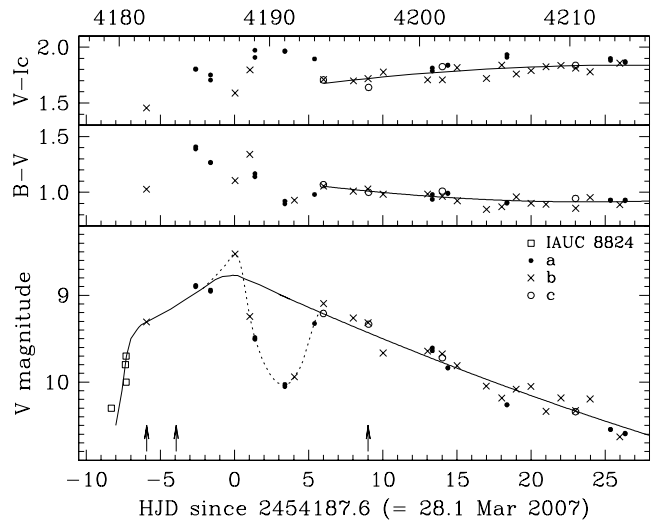}
     \caption{Enlargement of the early portion of Nova Oph 2007 photometric
            evolution. The meaning of the solid
            and dashed lines is outlined in Sect. 3.1. Symbols and
            ordinates as in Figure~1. The arrows point to times when
            we recorded optical spectra.}
   \end{figure}

\section{Properties of Nova Oph 2007}

\subsection{Rise, maximum brightness and early decline}

The early phase of NOph07 photometric evolution is shown in greater
detail in Figure~2.
The solid line in the $V$-band panel represents a smoothed interpolation of
the observed behavior, whose declining branch is given by the parabolic
expression $V$=$8.75 + 0.079\times t - 0.00045\times t^2$ where $t$ is the
time in days from maximum brightness. This expression well fits the observed
decline between $t$=+6 and $t$=+65 ($B$=$9.78 + 0.075\times t -
0.00043\times t^2$ would equally well fit the $B$-band light-curve over the
same period). On top of this smooth behavior, around maximum brightness
NOph07 displayed an oscillation (indicated by the dashed line in Figure~2)
that took $\sim$8 days to complete, and reached a peak-to-valley amplitude
$\Delta V$$\approx$1.3~mag.

The $V$-band maximum of NOph07 occurred around March 28.1, 2007 UT
(=2454187.6), when the nova was measured at $V$=8.52, $B-V$=+1.12,
$V-R_{\rm C}$=+0.76, $V-I_{\rm C}$=+1.59 and $R_{\rm C}-I_{\rm
C}$=+0.83. Decline times were $t^{V}_{2}$=26.5, $t^{B}_{2}$=30,
$t^{V}_{3}$=48.5 and $t^{B}_{3}$=56.5~days. They correspond to a 
moderately fast speed class.

\begin{figure*}
     \centering
     \includegraphics[height=15.0cm,angle=270]{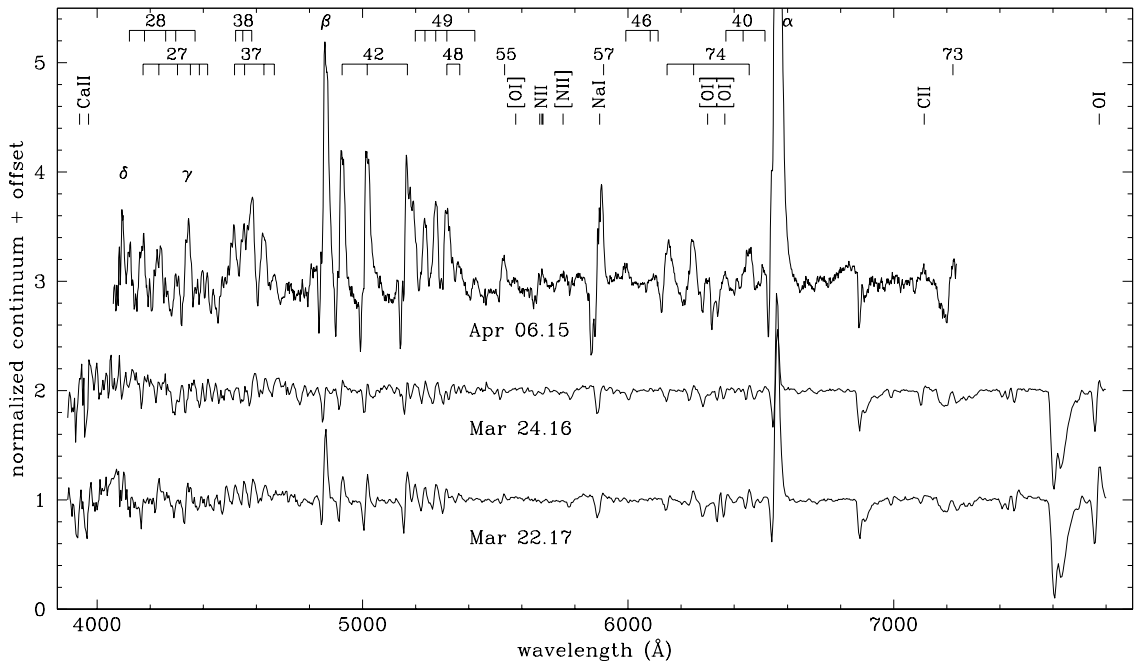}
   \caption{Spectral appearance of Nova Oph 2007 on 6 and 4 days before
            maximum and 9 days past it. The ordinate scale is the same 
            for all spectra, which are continuum normalized and off-set
            for better visibility.}
\end{figure*}

The color changes of NOph07 around maximum brightness have been quite large
and poorly correlated among the various bands (cf Figure~2), amounting to
$\Delta (B-V)$=0.51, $\Delta (V-R_{\rm C})$=0.20, $\Delta (V-I_{\rm C})$=0.38,
$\Delta (R_{\rm C}-I_{\rm C})$=0.19 mag. Once the oscillation noted around
maximum quenched down, the evolution of colors settled onto a smooth
behaviour similarly to that of $V$-band lightcurve. The parabolic fits given
as solid lines in Figure~2 correspdonds to $B-V= 1.165 - 0.0211\times t +
0.00044\times t^2$ and $V-I_{\rm C} = 1.566 + 0.0205\times t - 0.00038\times
t^2$ (between $t$=+6 and $t$=+65).

The initial rise of NOph07 to maximum was fast. Nakamura (2007) reports that
nothing was visible at nova position down to 13.0 mag at $t$=$-$10.3 days,
and down to 11.3 mag at $t$=$-$9.3 days according to Tago (2007). Combining
with discovery magnitude estimates reported in IAUC 8824 (cf Figure~2), this
corresponds to a rising rate {\it faster} than 1.2 mag~day$^{-1}$. At
$t$=$-$6.5, when the nova was $\Delta V$$\sim$0.9 mag below maximum, it
slowed its rising rate, entered a pre-maximum halt phase and completed 
the rise to maximum at a more leisurely 0.14 mag~day$^{-1}$ rate.

\subsection{Reddening}

van den Bergh and Younger (1987) derived a mean intrinsic color
$(B-V)_\circ$=+0.23 $\pm$0.06 for novae at maximum, and 
$(B-V)_\circ$=$-$0.02 $\pm$0.04 for novae at $t_2$. We measured for NOph07
$B-V$=+1.12 at maximum and $B-V$=+0.89 at $t_2$, which correspond 
respectively to $E_{B-V}$=0.89 and $E_{B-V}$=0.91. In the rest of this paper
we will adopt $E_{B-V}$=0.90 for NOph07 (corresponding to $A_V$=2.85 and
$A_B$=3.78 for a standard $R_V$=3.1 extinction law), which is in good
agreement with a preliminary $E_{B-V}$=1.0 reddening estimate from infrared
OI emission lines by Rudy et al. (2007).
The large reddening affecting NOph07 is confirmed by the large equivalent
width of interstellar NaI D lines that are easily visible on our low
resolution spectra superimposed on the emission component of NaI D P-Cyg
profile of the nova (but not visible at the compressed scale
of Figure~3). The resolution of these spectra is however too low to give
a reliable measure of the equivalent width of NaI D from which to derive
$E_{B-V}$ via the Munari and Zwitter (1997) calibration. From these
interstellar NaI D lines, we can only place a $E_{B-V}$$>$0.7 lower 
limit to the reddening.

\subsection{Distance}

The rate of decline from maximum and the observed magnitude 15 days past
maximum are calibrated tools to estimate distances to novae.

Published relations between absolute magnitude and rate of decline generally
take the form $M_{\rm max}\,=\,\alpha_n\,\log\, t_n \, + \, \beta_n$. Cohen
(1988) $M_V$-$t^V_2$ and Schmidt (1957) $M_V$-$t^V_3$ relations provide
$M_V$=$-$7.27 $M_V$=$-$7.29 for NOph07, respectively. Capaccioli et al.
(1989) and Schmidt-Kaler (1965, cf Duerbeck 1981) $M_B$-$t^B_2$ relation
give $M_B$=$-$7.24 and $M_B$=$-$7.46, respectively. de~Vaucouleurs (1978)
and Pfau (1976) $M_B$-$t^B_3$ relations lead to $M_B$=$-$7.10 and
$M_B$=$-$7.52, respectively, while della Valle and Livio (1995) $s$-shaped
relation calibrated on novae in LMC and M31 provides $M_V$=$-$7.56. Buscombe
and de Vaucouleurs (1955) suggested that all novae have the same absolute
magnitude 15 days after maximum light. Different calibrations of their
relation are available from Buscombe and de Vaucouleurs (1955), Schmidt
(1957), Pfau (1976), de Vaucouleurs (1978), Cohen (1985), van den Bergh and
Younger (1987), van den Bergh (1988), Capaccioli et al. (1989) and Downes
and Duerbeck (2000). The brightness of NOph07 15 days after maximum light,
derived from the parabolic fits in Figure~2, was $V_{\rm 15}$=9.83 and 
$B_{\rm 15}$=10.81 ($\pm$0.015).

The mean value for all these distance estimates (and its error of the mean)
is $d$=3.7 $\pm$0.2 kpc. At such distance the absolute magnitudes of NOph07
at maximum would have been $M_V^{\rm max}$=$-$7.2 and $M_B^{\rm
max}$=$-$7.0, and the height above the galactic plane is $z$=215~pc, which
is within the vertical scale height of the galactic Thin Disk.

\subsection{Dust formation ?}

Rudy et al. (2007) reported dust formation occuring in NOph07 during May
2007, when their infrared spectra of the nova were characterized by
persisting low ionization conditions, being dominated by FeII, NI, OI and CI
emission lines. According to Rudy et al., the dust was absent on their May~7
($t$=+50 days) observations and was instead substantially present on May~31
($t$=+74) when they estimated from OI emission lines the
reddening affecting the nova to have increased from $E_{B-V}$=1.0 to 1.3.

Dust generally forms in novae during the transition from stellar to nebular
spectra, right when the high ionization rapidly sweeps through the ejecta
(Gehrz et al. 1998, Shore and Gehrz 2004). Thus, the Rudy et al. (2007)
announcement of dust forming in NOph07 during persistent, low ionization
conditions would correspond to an unusual behaviour for a nova. The event
also has no counterpart on the optical lightcurve of Figure~1, where all
colors on May~31 are appreciably {\em bluer} than on May~7, contrary to
the reported increase by 0.3 in $E_{B-V}$.

  \begin{figure}
     \centering
     \includegraphics[width=5.5cm]{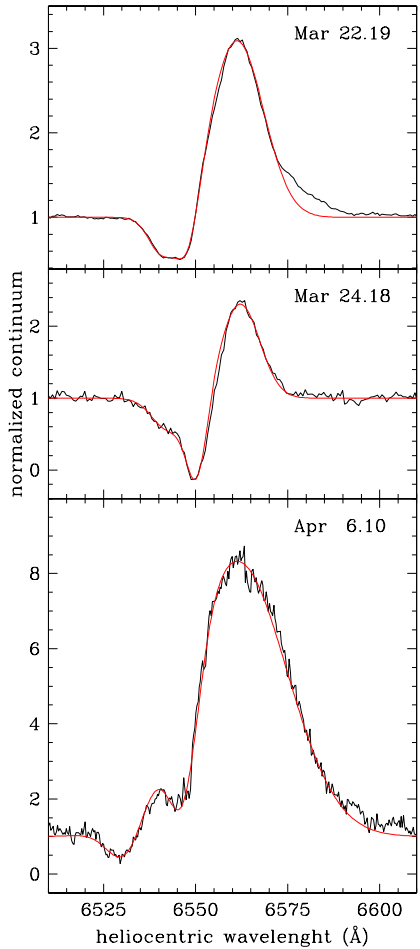}
   \caption{Continuum normalized H$\alpha$ profile of Nova Oph 2007 on 6 and 4 days
            before maximum and 9 days past it, at the same time of the
            low-res spectra of Figure~3. The fit with the emission and absorption
            components detailed in Table 2 is overplotted.}
  \end{figure}

  \setcounter{table}{1}
  \begin{table}
     \caption{Heliocentric velocity, velocity span at half maximum,
              equivalent width and integrated absolute flux (in
              erg cm$^{-2}$ sec$^{-1}$) of the emission and
              absorption components of the H$\alpha$ profiles shown in   
              Figure~4.}
     \centering
     \includegraphics[width=6.0cm]{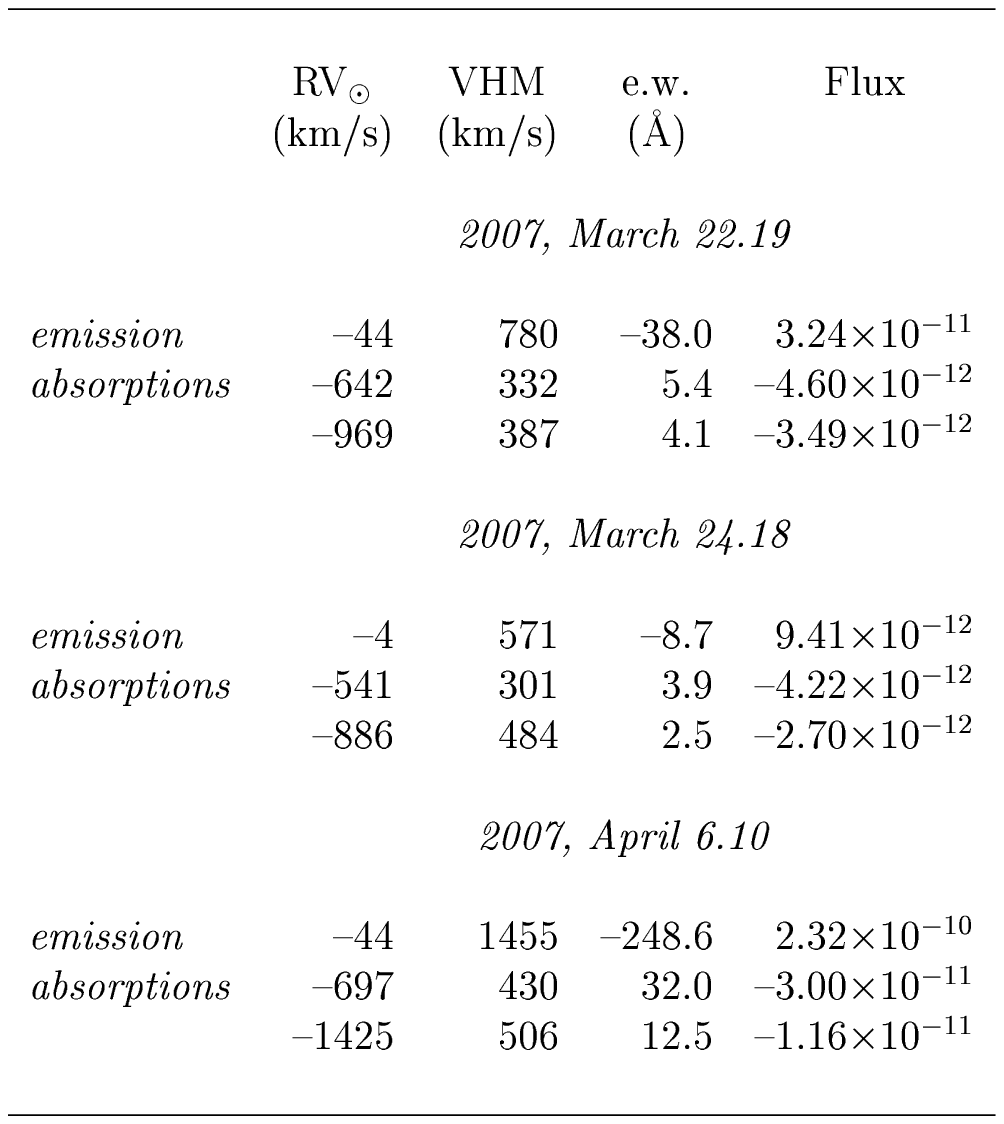}
     \label{tab2}
  \end{table}

Dust could have formed later, three months and $\Delta V$=5.5 mag
past maximum, peaking on June 22. Superimposed on a smooth light- and
color-evolution, the nova displayed between $t$$\approx$+82 and
$t$$\approx$+100 a fading and recovery in brightness, paralleled by
first reddening and then a return to normal optical colors. The event well documented by
Figure~1, amounted to $\Delta B$=1.8~mag, $\Delta (B-V)$=0.4 and $\Delta
(V-I_{\rm C})$=0.6, which nicely correspond to an increase in the reddening by
$\Delta E_{B-V}$=0.44. 

For a normal $R_V$=3.1 extiction law, a $\Delta E_{B-V}$=0.44 dust
condensation event should 
produce $\Delta (V-R_{\rm C})$=0.28 and $\Delta (R_{\rm C}-I_{\rm C})$=0.35
reddenings. We have instead observed $\Delta (V-R_{\rm C})$=0.5 and 
$\Delta (R_{\rm C}-I_{\rm C})$=0.1, that would correspond to 
$\Delta E_{B-V}$=0.80 and 0.12, respectively. Even if the mean
value would well agree with $\Delta E_{B-V}$=0.44 derived from $B$, $V$ and
$I_{\rm C}$ colors, nevertheless the observed $\Delta (V-R_{\rm C})$ and
$\Delta (R_{\rm C}-I_{\rm C})$ requires an explanation. To this aim, it is
sufficient that a fraction of flux in the H$\alpha$+[NII] 6458-84~\AA\
emission blend originates from gas external to the region where dust
condensed. This emission blend usually accounts for the vast majority of the
flux in the $R_{\rm C}$-band of novae during advanced decline. For sake of
discussion, if we assume that in NOph07 at the time of the $\Delta
E_{B-V}$=0.44 dust condensation event the H$\alpha$+[NII] blend was
contributing 80\% of the collected $R_{\rm C}$-band flux, to account for the
observed $\Delta (V-R_{\rm C})$=0.5 and $\Delta (R_{\rm C}-I_{\rm C})$=0.1
it is enough that 10\% of the H$\alpha$+[NII] 6458-84~\AA\ flux was not 
affected by the extinction. This could be easily the case because [NII]
originates in the most external parts of the ejecta, where the gas density
is possibly too low to support fast and efficient dust grain formation.

To be properly addressed, the issue as to whether dust actually formed in Nova Oph
2007, when and how much, and its radial location within the expanding
ejecta, will have to wait for the publication of all available information,
expecially infrared and spectroscopic data.

\subsection{Spectral evolution}

  \begin{figure}
     \centering
     \includegraphics[width=6.0cm]{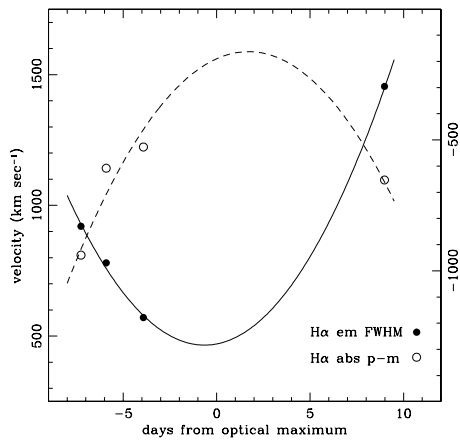}
   \caption{Evolution around maximum brightness of the FWHM of the
            emission component of H$\alpha$ profile (ordinate scale at left) and 
            of the velocity of the pre-maximum absorption system (ordinate scale at
            right). Data from Table~2 and Naito and Narusawa (2007) for
            $t$=$-$7.3. The fits are explained in Sect. 3.5.}
  \end{figure}

The evolution of low-resolution spectra of NOph07 around maximum brightness
is presented in Figure~3, where all significant emission lines are
identified. We obtained the first spectrum at $t$=$-$6. It is a charaterized
by low ionization conditions and weak emission lines of mainly FeII, Balmer
and OI, flanked by P-Cyg absorption profiles. The overall intensity of
absorption lines, in particular of CH~4310~\AA, CaII H and K, NaI D and
H$\gamma$ support a classification as an F2-3 supergiant (allowing for a slight
overabundance of Carbon in the ejecta, as typical of novae). This matches
the observed $B-V$=+1.0 color on the same UT date. Two days later and 0.2
mag closer to maximum brightness on $t$=$-$4, all emission lines had
weakened considerably, with only H$\alpha$, [OI]~7772~\AA\ and FeII 4923,
5018~\AA\ lines still displaying a detectable emission component. At the
same time the underlying absorption spectrum increased, following a pattern
quite typical for novae (cf McLaughlin 1960, hereafter M60) that sees a
decrease in the ionization conditions and cooling of the spectral continuum
along the rise to maximum brightness, and a reversed pattern durind the
decline from it. The intensity of the absorption lines suggest an F5 supergiant
to be the closest spectral classification.  Remarkably strong diffuse
interstellar bands (in particular 4430, 5780, 6284 and 6614~\AA) and
interstellar lines (NaI D and CaII H and K) rise above the continuum of
Figure~3. The spectrum for $t$=+9 and $\Delta V$=0.9 mag down from maximum
is that of a classical FeII-class nova, with all relevant FeII
multiplets in emission, Balmer and NaI lines also in strong emission, and
feeble traces of [OI], CII, [NII] and NII just beginning to emerge.

Figure~4 presents the evolution of the H$\alpha$ profile from high resolution
observations obtained on the same dates of the low resolution spectra of
Figure~3. All three profiles clearly indicate the presence of two absorption
components in addition to the emission one. A fitting with three gaussian
components is overplotted to the observed profiles in Figure~4, and their
radial velocity, FWHM, equivalent width and absolute flux are listed in
Table~2. The post maximum spectrum on $t$=+9 is characterized by an
expansion velocity of the ejecta (estimated from the H$\alpha$ emission
component) of 730~km~sec$^{-1}$, and the presence of the principal and
diffuse enhanced absorption systems, whose displacement from the emission
component are  $\sim$650 and 1380 km/s, respectively. The statistical
relations by M60 would predict for the $t_2$, $t_3$ decline rates 
of NOph07 a velocity for these absorption systems of 700 and 1350~km/s,
pretty close to observed values.

Figure~5 illustrates the evolution with time of the FWHM of the H$\alpha$
emission component and of the radial velocity of the principal absorption
system, combining data in Table~2 with Naito and Narusawa (2007) earlier 
measurement for $t$=$-$7.3. Both suggest an evolution that reaches minimum
values around the time of maximum brightness, as observed in other novae (cf. M60).
For sake of documentation and without attaching to them excessive
significance given the limited number of observational points they rest
upon, the parabolic fitting in Figure~5 of the  FWHM of the H$\alpha$
emission component is given by FWHM=470 + 13.8$\times t + 10.6 \times t^2$,
and that for the velocity of the principal absorption system is
RadVel = $-$190 + 32$\times t - 9.4 \times t^2$.

\subsection{Late photometric evolution}

The photometric evolution presented in Figure~1 is characterized by a marked
flattening, settling in around $t$=+130, that interrupted the normal decline
when the nova was $\Delta V$=7.0~mag fainter than maximum. The effect is
real because ($i$) it is present in data collected independently with
different instruments, and ($ii$) there is no evidence for an optical faint
companion neither in our images, nor in DSS, 2MASS or DENIS survey data down
to $V\approx$20 mag, which could have perturbed the measurement of NOph07.
The approaching conjunction with the Sun stopped our monitoring at $t$=+200
and with it the possibility to further follow this interesting photometric 
phase.

\bsp

\label{lastpage}

\end{document}